\documentstyle[12pt]{article}
\begin{document}
 
\title {\bf Order in Binary Sequences \\
and the Routes to Chaos}
\author{Ricardo L\'{o}pez-Ruiz  \\
                                   \\
{\small Department of Computer Science and BIFI,} \\
{\small Facultad de Ciencias, Edificio B,}\\
{\small Universidad de Zaragoza,} \\
{\small 50009 - Zaragoza (Spain)}}

\maketitle
\baselineskip 8mm
 
\begin{center} {\bf Abstract} \end{center}
The natural order in the space of binary sequences
permits to recover the $U$-sequence. Also the scaling
laws of the period-doubling cascade and the intermittency route to chaos
defined in that ordered set are explained. 
These arise as intrinsic properties of this ordered set, 
and independent from any consideration about dynamical systems.


\newpage

\section{Introduction}

In the last decades many efforts have been devoted to understand and 
to describe the different routes to chaos that are found in
the dynamical behavior of the most diverse physical systems. 
Many of these situations can be modeled by unimodal maps, 
which form a family of well understood dynamical systems \cite{collet}.
Some characteristics of the unimodal maps are 
gathered in three basic works \cite{metropolis,feigenbaum,pomeau}.
Metropolis {\it et al.} \cite{metropolis}
established the order in which the periodic 
orbits appear in that kind of systems. They found the {\it $U$-sequence},
which is a universal sequence of bifurcations common to all unimodal maps. 
Feigenbaum \cite{feigenbaum} discovered the constant that brings his name
and that is a consequence of the universal scaling properties 
of the period-doubling cascade. 
Pomeau and Manneville \cite{pomeau} proposed simple models to understand 
and to catalog some dynamical mechanisms of bursting
in temporal signals.

The fact that so simple systems display so many universal 
properties is remarkable. In this small piece of work,
it is our aim to offer a unified view of a similar panorama 
happening in the space of the binary sequences. 
For this purpose, the natural order in the space of  
binary sequences will be recalled in section 2. 
Besides the recovering of the $U$-sequence, 
the scaling laws for alternative period-doubling cascade 
and intermittency 
routes to chaos defined in this binary scheme will be found.
These laws are intrinsic properties of the ordered set
of binary sequences, and they are also independent from
any consideration about dynamical systems.
This optional and simplified view will be presented in section 3. 
The last section contains the conclusions.

\section{Order in the Space of Binary Sequences} 

Let us recall the nomenclature introduced in Ref. \cite{lopezruiz}.
{\it $\cal B_S$} represents the space of binary sequences. 
An orbit of period $n$ ($O_n$) is formed by a binary sequence 
non periodic (irreducible) of $n$ digits and all sequences built 
from the iteration of this irreducible sequence as a block. 
It is represented by the irreducible orbit. An example can be: 
$O_3=100\equiv \{100,100100,100100100,\ldots \}$. The {\it real equivalent}
of the orbit $O_n=\alpha_{n-1}\alpha_{n-2}\ldots\alpha_1\alpha_0$ is 
the real number $\sum_{i=0}^{n-1}\alpha_i2^i$. The set formed by
an $n$-periodic orbit, $O_n$, and its $n-1$ cyclic permutations will be called
{\it orbital} of period $n$ ($[O_n]$). For example, 
$[O_3]=[100]\equiv \{100,010,001\}$. The set of the conjugate orbits is the
{\it conjugate orbital}: $[\overline{O}_3]=[011]\equiv \{011,101,110\}$.
The number of orbitals $N_n$  of period $n$ is given by:
\begin{equation} 
N_n=\frac{2^n-\sum_{i=1}^{k}m_iN_{m_i}}{n},
\end{equation}
where $\{m_1,m_2,\ldots,m_k\}$ are the integer divisors of $n$ 
excluding $\{n\}$. \par

{\bf \underline{Lorenz Order}:}
If $[O_n]\neq [\overline{O}_n]$, the set $L_n^D=[O_n;\overline{O}_n]\equiv
[O_n]\cup [\overline{O}_n]$ is called {\it $L$-doublet}.
If $[O_n]=[\overline{O}_n]$, the set $L_n^S=[O_n;]\equiv [O_n]$ 
is called {\it $L$-singlet}. Given a $L$-doublet 
or $L$-singlet, the orbit that starts in the left side by $1$  followed
by the subsequence with the smallest real equivalent is called the
{\it characteristic orbit} ($O_n^c$). This orbit is chosen as representative:
$L_n^D=[O_n^c,\overline{O}_n^c]$, $L_n^S=[O_n^c;]$. For example,
if $O_3^c=100$ then $L_3^D=[100;011]$ or if $O_4^c=1001$ then $L_4^S=[1001;]$.
The {\it associated fraction} ($r$) of a $L$-doublet or a $L$-singlet 
is a fraction associated to its characteristic orbit,
$O_n^c=\alpha_1\alpha_2\ldots\alpha_n$ and defined as:
\begin{equation} 
r=\left(\sum_{i=1}^{n}\frac{\alpha_i}{2^i}\right)\cdot\frac{2^n}{2^n-1}.
\end{equation}
These fractions are (except the trivial $1$-periodic 
orbit with $r_{[1;0]}=1$) in the range 
$r_{[10;]}=\frac{2}{3}>r>\frac{1}{2}=r_{[100\stackrel{\infty}{\ldots};
011\stackrel{\infty}{\ldots}]}$. 

Now we will establish an order relation
in the space of $L$-doublets and $L$-singlets,
called {\it implication},
and represented by the symbol $\Rightarrow$.
A generic element of this space is represented by the symbol $[;]$. 
We will said that the element $[;]_i$ implies the element $[;]_j$ 
when the associated fractions $(r_i,r_j)$ verify $r_i<r_j$:
\begin{equation}
[;]_i \Rightarrow [;]_j \leftrightarrow r_i<r_j.
\end{equation}
For example, $L_4^D\equiv [1000;0111]\Rightarrow L_3^D\equiv [100;011]$
because $r_{[1000;0111]}=\frac{8}{15}<\frac{4}{7}=r_{[100;011]}$.
By applying these rules the ordered binary set showed in table 1
is found. Each element in the table implies all elements above it. 
We represent the space of binary sequences and the Lorenz-order relationship
by $({\cal B_S}, {\cal L_{\Rightarrow}})$. 

{\bf \underline{R\"ossler Order} ($U$-sequence):} 
Another order type, which we call R\"ossler order, is established in $\cal B_S$.
This order generates the $U$-sequence. It is built from
the Lorenz-order with the following steps.
We make a different orbit grouping in $({\cal B_S},{\cal L_{\Rightarrow}})$: 
each $L$-doublet, $L_n^D=[O_n^c;\overline{O}_n^c]$, and 
its doubled $L$-singlet, $L_{2n}^S=[O_n^c\overline{O}_n^c;]$, 
is grouped in a {\it $L$-triplet}, 
$L_n^T=[O_n^c,\overline{O}_n^c;O_n^c\overline{O}_n^c]=L_n^D\cup L_{2n}^S$.
An $L$-singlet not belonging to any $L$-triplet is called {\it $L$-singlet$_d$}.
They have the property: $L_{2n}^{S_d}=[O_n\overline{O}_n;]\rightarrow 
O_n=O_{\frac{n}{2}}\overline{O}_{\frac{n}{2}}$. Then, $\cal B_S$ is 
divided in a new partition: $L$-triplets and $L$-singlets$_d$. 
We define the binary sequence transformation $\cal F_{B_S}$ 
that transforms the binary sequence
$\{l_1l_2\ldots l_n\}$ in the binary sequence 
$\{r_1r_2\ldots r_n\}$ according to the law:
\begin{eqnarray}
r_{i} & = & l_i+l_{i+1}\pmod{2} \\
l_{n+1} & = & l_1.  \nonumber
\end{eqnarray}
The inverse transformation $\cal F_{B_S}$$^{-1}$ is defined as follows:
\begin{equation}
l_i=\sum_{j=0}^{i-1} r_j,
\end{equation}
with $r_0=1$ and $r_{n+i}=r_i$. This last transformation must be applied
to the doubled sequence $\{r_1r_2\ldots r_nr_1r_2\ldots r_n\}$ in order to
get the doubled sequence $\{l_1l_2\ldots l_nl_1l_2\ldots l_n\}$.

If we apply $\cal F_{B_S}$ to the characteristic orbit $O_n^c$ of a $L$-triplet
the result is the regular-orbit $O_n^r$, an orbit with even number of $1$s.
If $\cal F_{B_S}$ is applied to the orbit $O_n^c\overline{O}_n^c$ 
of the same $L$-triplet the result is the flip-orbit $O_n^f$, 
an orbit with odd number of $1$s. Thus the $L$-triplet is transformed by
$\cal F_{B_S}$ into two independent orbitals whose union, 
$R_n^D=[O_n^r;O_n^f]\equiv
[O_n^r]\cup [O_n^f]$, is called an {\it $R$-doublet}.
The transformation of the $L$-singlet$_d$ $L_{2n}^{S_d}$ by $\cal F_{B_S}$ 
produce an orbital called an {\it R-singlet}, $R_n^S$. The action of
$F_{S_B}$ is summarized as follows:
\begin{eqnarray*}
{\bf \cal F_{B_S}}:\;\;\;\; {\cal B_S} & \longrightarrow & {\cal B_S} \\
L_n^T\left\{\begin{array}{c}
L_n^D \\
L_n^S
\end{array}\right.          & 
\begin{array}{c}
\longrightarrow \\
\longrightarrow 
\end{array}                 & 
\left.\begin{array}{c}
[O_n^r] \\
\,[O_n^f]
\end{array}\right\} R_n^D   \\
L_{2n}^{S_d} & \longrightarrow & R_n^S
\end{eqnarray*}
Moreover, $\cal F_{B_S}$ transfers the Lorenz-order in 
($\cal B_S$, $\cal L_{\Rightarrow}$) to a new order called R-order
(Table 2).
This order relationship is defined as follows: if 
$L_{n_i}^{S_d,T}\Rightarrow L_{n_j}^{S_d,T}$ then the transformed orbitals
by $\cal F_{B_S}$ verify that
$R_{n_i}^{D,S}\Rightarrow R_{n_j}^{D,S}$. (This is well-defined because
the L-doublet and the L-singlet comprising the L-triplet do not have
any other orbital between them in the L-order).
The new ordered space is denoted by ($\cal B_S$, $\cal R_{\Rightarrow}$).
This is the $U$-sequence (see Table 1).

\section{Routes to Chaos}
 
An infinite subset $\cal C$ of the ordered set 
$({\cal B_S}, {\cal L_{\Rightarrow}})$ is called a {\it route to chaos} 
if $([;]_i,[;]_j)\in {\cal C}$ and $[;]_i\Rightarrow [;]_k\Rightarrow [;]_j$ 
implies that $[;]_k\in {\cal C}$. There are in the ordered set $({\cal B_S}, {\cal L_{\Rightarrow}})$
two important kinds of route to chaos with their respective scaling laws 
that we proceed to present now.

{\underline{\it Period-doubling route to chaos}, $\cal C_{PD}$}: 
This set is formed by a $L$-doublet, $L_n^D=[O_n^c;\overline{O}_n^c]$, 
and all the consecutive $L$-singlets
of double period. That is ${\cal C_{PD}}\equiv
\{L_n^D,L_{2n}^S,L_{4n}^S,\ldots,L_{\infty}^S\}$. For example,
$\{1,10,1001,10010110,\ldots\}$. 

Let us calculate the sequence of
associated fractions $\{r_n,r_{2n},r_{4n},\ldots,r_{\infty}\}$. 
A straightforward calculation gives us:
\begin{equation}
r_{2n}=\left[ r_n(2^n-1)+1\right]\frac{2^n-1}{2^{2n}-1}.
\end{equation}
The scaling law associated to this route to chaos is:
\begin{equation}
\lim_{n\rightarrow \infty}\frac{2^{2n}(r_n-r_{2n})}{2^{4n}(r_{2n}-r_{4n})}=1.
\end{equation}
This law is independent of the particular set $\cal C_{PD}$ and it is 
intrinsic to the own structure of the ordered space of binary sequences.
(This is reflected in the Feigenbaum constant for unimodal maps with quadratic 
critical point).

{\underline{\it Intermittency route to chaos}, $\cal C_{IT}$ }: 
Let us take any $L$-doublet
expressed by $L_p^D=[O_p;\overline{O}_p]$ where 
$[O_p]=[\alpha_1\alpha_2\ldots\alpha_p]$. We define the orbital
$[O_{p+1}]=[\alpha_1\alpha_2\ldots\alpha_p1]$ and the $L$-doublet
associated to it, i.e., $L_{p+1}^D=[O_{p+1},\overline{O}_{p+1}]$. The route to chaos 
$\cal C_{IT}$ defined by these two extrema $L$-doublets:
${\cal C_{IT}}\equiv\{L_{p+1}^D,\ldots,L_{p}^D\}$ 
is called the intermittency route to chaos. An example can be:
$\{[1001;],\ldots,[100;011]\}$. We choose a subsequence of $\cal C_{IT}$
in the following form:
\begin{eqnarray*}
L_{p+1}^D=[O_{p+1},\overline{O}_{p+1}] & \rightarrow &
O_{p+1}=\alpha_1\ldots\alpha_p1 \\
L_{2p+1}^D=[O_{2p+1},\overline{O}_{2p+1}] & \rightarrow &
O_{2p+1}=\alpha_1\ldots\alpha_p\alpha_1\ldots\alpha_p1 \\
\vdots & \vdots & \vdots \\
L_{np+1}^D=[O_{np+1},\overline{O}_{np+1}] & \rightarrow &
O_{np+1}=\alpha_1\ldots\alpha_p\stackrel{(n)}{\cdots} \alpha_1\ldots\alpha_p1\\
\vdots & \vdots & \vdots \\
L_{p}^D=[O_{p},\overline{O}_{p}] & \rightarrow &
O_{p}=\alpha_1\ldots\alpha_p 
\end{eqnarray*}
The subsequence from the last example is: 
$\{1001,1001001,1001001001,\\ \ldots,100100\stackrel{100}{\ldots}1001,
\ldots,100\}$. Given the fraction $r_p$ associated to $L_p^D$, 
a straightforward calculation permits to find the associated 
fractions of the remaining fraction subsequence:
\begin{equation} 
r_{np+1}=\frac{2r_p(2^{np}-1)+1}{2^{np+1}-1}.
\end{equation}
Then, the subsequence $\{r_{p+1},r_{2p+1},\ldots,r_{np+1},\ldots,r_p\}$
is obtained. We denote by $\delta=r_{np+1}-r_p$ the distance to the critical
fraction $r_p$ where the periodic behaviour take place and
the integer $l=np$ of an orbit $O_{np+1}$ represents the binary length 
of the periodic sequence of this orbit ('phase laminar' length). 
The isolated digit $1$ that breaks this periodic behavior is called
a 'burst'. If we calculate the relation between $\delta$ and $l$,
the scaling law for the intermittency route to chaos is obtained:
\begin{equation}
r_{np+1}-r_p=\delta\sim\frac{1-r_p}{2^{np+1}}\Longrightarrow
4\delta\cdot 2^l\sim 1,
\end{equation}
where $r_p$ has been taken as $0.5$. Let us observe that
when the distance to the critical point goes to zero, we recover 
the divergence of laminar phase length as an intrinsic and universal 
property of the ordered set of binary sequences.

\section{Conclusions}

The properties of the routes to chaos in high 
dimensional systems is still a not very well understood subject \cite{fournier}. 
Unidimensional results are generalized for instance in the work \cite{lopez},
where an orbit implication diagram for horseshoe type-flows is calculated 
by topological methods. This provides a partial order on orbit formation  
for three dimensional flows and two dimensional orientation preserving 
maps which evolve into horseshoe under parameter variation. The results are
independent of dissipation from the conservative limit (zero dissipation)
to the unimodal limit (infinite dissipation).

In this work, two routes to chaos have been defined in 
the space of binary sequences. One of them mimics the 
period-doubling cascade arising in the unimodal maps 
and the other one mimics the intermittency route to chaos.
The scaling properties
for these both routes to chaos have been calculated.
A Feigenbaum-like relationship for the bifurcation parameters
in the period doubling case and the dependence
of the laminar phases length on the distance to the 
critical point in the intermittency scenario seem to be
a consequence of the intrinsic properties of the
ordered binary set, and independent from any consideration
about dynamical systems. This simple and primitive view of
the route to chaos could be interpreted as the {\it radiograph}
of these dynamical phenomena when they are observed in more 
complicated systems.

{\bf Acknowledgements:}
This problem comes from Philadelphia (1993).
I must thank Prof. R. Gilmore (Philadelphia) and 
Prof. G. Mindlin (Buenos Aires) for very interesting discussions.

\newpage

\begin{center} Table Captions \end{center}

{\bf Table 1.} R\"ossler-order ({$\cal R_{\Rightarrow}$}) 
and Lorenz-order ({$\cal L_{\Rightarrow}$})
in the set of binary sequences until
period $n=6$. Also the order of the orbits 
in unimodal maps is given. ($^*$ means period-doubled orbits).

{\bf Table 2.} The two different orbit groupings, R\"ossler and Lorenz type, 
established in the space of binary sequences.

\newpage
\begin{table}[htb]
\begin{center}
\begin{tabular}{|l|c|l|c|c|} \hline
$R-Order$  & $Period_{order}$  & $L-Order$ & $r$ & $r$ \\ \hline\hline 

$0      $   & $ 1_1  $  & $1$ ; $0$          & $1       $& $1.00000$  \\ 
$1      $   &           & $10          $     & $2/3     $& $0.66666$  \\ \hline
$10     $   & $ 2_1^*$  & $1001        $     & $3/5     $& $0.60000$  \\ \hline
$1011   $   & $ 4_1^*$  & $10010110    $     & $10/17   $& $0.58823$  \\ \hline
$101110 $   & $ 6_1  $  &$100101$ ; $011010$ & $37/63   $& $0.58730$  \\ 
$101111 $   &           & $10010101101 $     &$2394/4095$& $0.58461$  \\ \hline
$10111  $   & $ 5_1  $  & $10010$ ; $01101$  & $18/31   $& $0.58064$  \\ 
$10110  $   &           & $1001001101$       & $589/1023$& $0.57575$  \\ \hline
$101    $   & $ 3_1  $  & $100$ ; $011$      & $4/7     $& $0.57142$  \\ 
$100    $   &           & $100011$           & $35/63   $& $0.55555$  \\ \hline
$100101 $   & $ 6_2^*$  & $100011011100$     &$2268/4095$& $0.55384$  \\ \hline
$10010  $   & $ 5_2  $  & $10001$ ; $01110$  & $17/31   $& $0.54838$  \\ 
$10011  $   &           & $1000101110$       & $558/1023$& $0.54545$  \\ \hline
$100111 $   & $ 6_3  $  & $100010$ ; $011101$& $34/63   $& $0.53968$  \\ 
$100110 $   &           & $100010011101$     &$2205/4095$& $0.53846$  \\ \hline
$1001   $   & $ 4_2  $  & $1000$ ; $0111$    & $8/15    $& $0.53333$  \\ 
$1000   $   &           & $10000111$         & $135/63  $& $0.52941$  \\ \hline
$100010 $   & $ 6_4  $  & $100001$ ; $011110$& $11/21   $& $0.52380$  \\ 
$100011 $   &           & $100001011110$     &$2142/4095$& $0.52307$  \\ \hline
$10001  $   & $ 5_3  $  & $10000$ ; $01111$  & $16/31   $& $0.51612$  \\ 
$10000  $   &           & $1000001111$       & $527/1023$& $0.51515$  \\ \hline
$100001 $   & $ 6_5  $  & $100000$ ; $011111$& $32/63   $& $0.50793$  \\ 
$100000 $   &           & $100000011111$     &$2079/4095$& $0.50769$  \\ \hline
\end{tabular}
\caption{}
\end{center}
\end{table}

\begin{table}[htb]
\begin{center}
\begin{tabular}{|c|c|} \hline
$Lorenz$ $Grouping$           &  $R\ddot{o}ssler$ $Grouping$\\ \hline\hline 

Binary                             &    Binary               \\
Sequences                          &    Sequences            \\ \hline
Orbits, $O_n$                      &    Orbits, $O_n$        \\
Orbitals, $[O_n]$                  &    Orbitals, $[O_n]$    \\ \hline
$L$-doublet, $L_n^D$               &                         \\
$L$-singlet, $L_n^S$               &                         \\ \hline
$L$-triplet, $L_n^T$               &    $R$-doublet, $R_n^D$ \\
$L$-singlet$_d$, $L_{2n}^{S_d}$    &    $R$-singlet, $R_n^S$ \\ \hline
\end{tabular}
\caption{}
\end{center}
\end{table}


\newpage
\begin{thebibliography}{16}

	 \bibitem{collet}
P. Collet and J.P. Eckmann, {\it Iterated maps on the interval as dynamical systems},
Birkha\"user, Boston (1980).

     \bibitem{metropolis}
N. Metropolis, M.L. Stein and P.R. Stein,
{\it On finite limit sets for transformations on the unit interval},
J. Combinatorial Theory {\bf 15}:25-44 (1973).

     \bibitem{feigenbaum}
M.J. Feigenbaum,
{\it Quantitative universality for a class of nonlinear transformations},
J. Statist. Phys. {\bf 19}:25-52 (1978).

     \bibitem{pomeau}
Y. Pomeau and P. Manneville,
{\it Intermittent transition to turbulence in dissipative dynamical systems},
Commun. Math. Phys. {\bf 74}:189-197 (1980).

	\bibitem{lopezruiz}
R. L\'opez-Ruiz, {\it A binary approach to the Lorenz model},
Chaos, Solitons and Fractals {\bf 8}, 1-6 (1997). 

	\bibitem{fournier}
D. Fournier-Prunaret, R. L\'opez-Ruiz and A. Taha, 
{\it Route to chaos in three-dimensional maps of logistic map}, 
ECIT'04-ITERATION THEORY, preprint nlin.CD/0502012 (2005).

     \bibitem{lopez}
G.B. Mindlin, R. L\'opez-Ruiz, H.G. Solari and R. Gilmore,
{\it Horseshoe implications},
Phys. Rev. E {\bf 48}:4297-4304 (1993).

\end{thebibliography}
\end{document}